\documentclass{aa}
\usepackage{booktabs,aalongtable}
\usepackage{natbib}
\usepackage{txfonts}
\usepackage{graphicx}
\bibpunct{(}{)}{;}{a}{}{,}
\def\fermi{\textit{Fermi~}}
\def\fermilat{\textit{Fermi}/LAT~}
\usepackage{lscape}
\usepackage{subfigure}
\usepackage{array}

\begin{document}

\title{Why are some BL Lacs detected by \fermi, but others not ?}
\author{}
\author{Zhongzu Wu \thanks{e-mail: zzwu08@gmail.com} \inst{1}, D. R. Jiang
\inst{2}, Minfeng Gu \inst{2}, Liang Chen \inst{2}
       }

\institute{College of Science, Guizhou University, Guiyang 550025, China.
        \and
        Key Laboratory for Research in Galaxies and Cosmology,
Shanghai Astronomical Observatory, Chinese Academy of Sciences, 80
Nandan Road, Shanghai 200030, China
        }

\date{Received / Accepted}
\abstract{
By cross-correlating an archival sample of 170 BL Lacs with 2 year \fermilat AGN sample, we have compiled a sample of 100 BL Lacs with \fermi detection (FBLs), and a sample of 70 non-\fermi BL Lacs (NFBLs). We compared various parameters of FBLs with those of NFBLs, including the redshift, the low frequency radio luminosity at 408 MHz ($L_{\rm 408MHz}$), the absolute magnitude of host galaxies ($M_{\rm host}$), the polarization fraction from NVSS survey ($P_{\rm NVSS}$), the observed arcsecond scale radio core flux at 5 GHz ($F_{\rm core}$) and jet Doppler factor; all the parameters are directly \textbf{measured} or derived from available data from literatures.
We found that the Doppler factor is on average larger in FBLs than in NFBLs, and the $Fermi~ \gamma$-ray detection rate is higher in sources with higher Doppler factor. In contrast, there are no significant differences in terms of the intrinsic parameters of redshift, $ L_{\rm 408MHz}$, $ M_{\rm host}$ and $ P_{\rm NVSS}$.  FBLs seem to have a higher probability of exhibiting measurable proper motion. These results strongly indicate a higher beaming effect in FBLs compared to NFBLs. The radio core flux is found to be strongly correlated with $\gamma$-ray flux, which remains after excluding the common dependence of the Doppler factor. At the fixed Doppler factor, FBLs have systematically larger radio core flux than NFBLs, implying lower $\gamma$-ray emission in NFBLs since the radio and $\gamma$-ray flux are significantly correlated. Our results indicate that the Doppler factor is an important parameter of $\gamma$-ray detection, the non-detection of $\gamma$-ray emission in NFBLs is likely due to low beaming effect, and/or low intrinsic $\gamma$-ray flux, and the gamma-rays are likely produced co-spatially with the arcsecond-scale radio core radiation and mainly through the SSC process.

 }

\authorrunning{Z. Z. Wu et al.}
\titlerunning{\fermi and non-\fermi BL Lac objects}

\maketitle

\section{Introduction}

Blazars are an extreme subclass of AGNs with characteristic properties such as
rapid variability at all wavelengths, high optical polarization, apparent superluminal motion,
flat radio spectra,  and a broad
continuum extending from the radio through the $\gamma$-rays \citep{urry95}.  The distinctive characteristic of blazars is a relativistic
jet oriented close to our line-of-sight.  BL Lac objects are blazars with  only very weak or non-existent emission lines
(equivalent width $< 5 ~\rm \AA$; e.g., Scarpa \& Falomo 1997).

They can be classified as different subclasses based on the synchrotron peak of their spectral
energy distribution (SED), namely, low-frequency peaked BL Lac
objects (LBL), intermediate objects (IBL) and high frequency peaked
BL Lac objects (HBL) \citep{padovani95}. The boundaries used for this work are as follows:
for LBLs, log\,$\nu_{\rm peak}<$\ 14.5, for IBLs 14.5 $<$ log\,$\nu_{\rm peak}<$\  16.5,
and for HBLs log\,$\nu_{\rm peak}>$\ 16.5\citep{nieppola2006}. The Large Area Telescope (Atwood et al. 2009, LAT) on board the \fermi $\gamma$-ray
Space Telescope  has been scanning the entire $\gamma$-ray sky approximately once every three hours
since July of 2008.  The LAT AGN catalogs \citep{abdo2009,abdo2010,ackermann2011} have shown that
BL Lac objects have been the most numerous group of $\gamma$-ray sources. The most recent LAT AGN catalog is the LAT Second Catalog of AGN (2LAC; Ackermann et al. 2011), the Clean Sample(sources with single associations and
not affected by analysis issues) includes 310 FSRQs, 395 BL Lacs, 24 other AGNs, and 157 of unknown type, which makes it possible to study the statistical properties of $\gamma$-ray sample of BL Lac objcets.

\begin{figure}
\centering
\includegraphics[height=5.5cm]{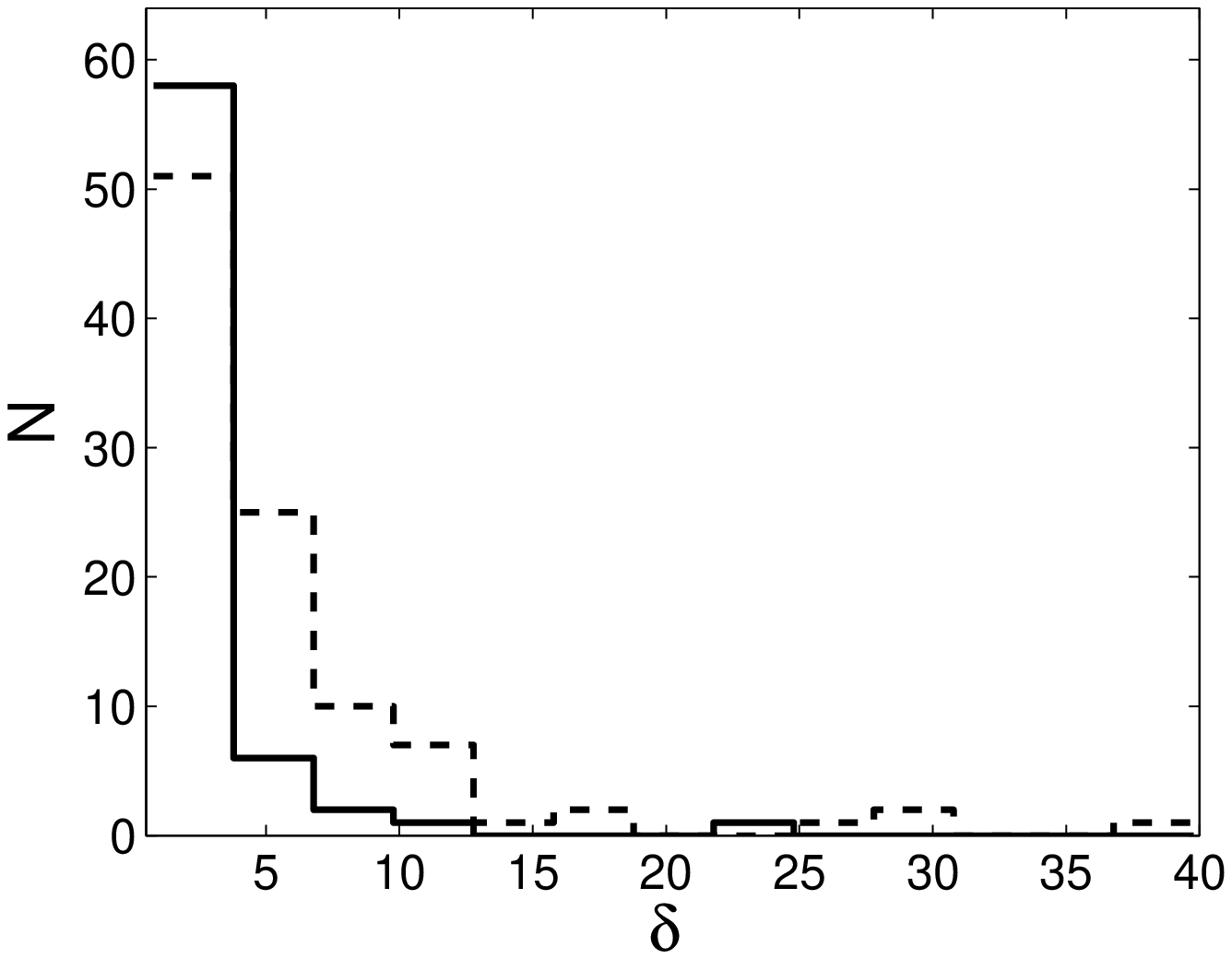}
\includegraphics[height=5.5cm]{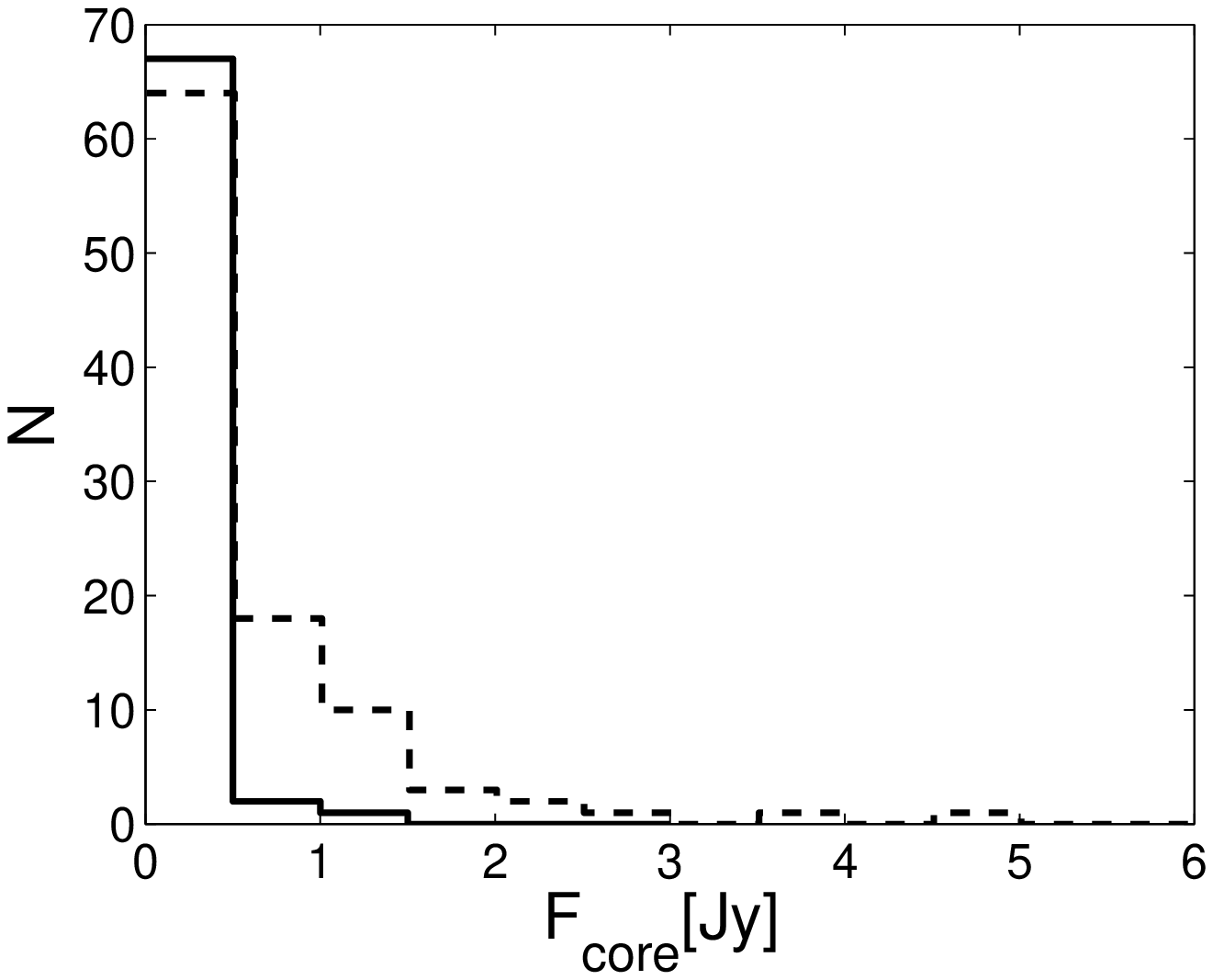}
\caption{The distributions of Doppler factor $\delta$ (top) and radio core flux $F_{\rm core}$ (bottom) for our BL Lacs. The dashed lines are for FBLs, while the solid lines for NFBLs.}
\label{dop_dis}
\end{figure}

By now, several possible answers have been proposed to the question ``why are some sources $\gamma$-ray loud and others are $\gamma$-ray quiet ?".
Doppler boosting is believed to be one of the important answers for this question. \cite{pin12} showed that sources in the \fermi LAT Second Source Catalog (2FGL; Nolan et al. 2012) display higher apparent speeds than those that have not been detected. \cite{pushkarev12aa} also showed that the \fermi AGNs have higher VLBI core flux densities and brightness temperatures, and are characterized by the less steep radio spectrum of the optically thin jet emission.
\cite{linford11,linford12} showed that the LAT flat-spectrum radio quasars (FSRQs) are
significantly different from the non-LAT FSRQs, while the \fermi detected BL Lacs(FBLs) tend to be generally similar to the non-\fermi BL Lacs(NFBLs); for example, there are no differences in the fraction of polarized BL Lac objects or the distributions of the polarization.  Irrespective of the general similarity, some differences between FBLs and NFBLs have been found. \cite{linford11,linford12} reported that the FBLs have longer jets, and are polarized more often, but core polarization itself seems not enough to separate two populations. Alternatively, the NFBLs may enter a state where $\gamma$-ray production ceases or is at least significantly reduced. However, the sample size of NFBLs in \cite{linford12} is rather small compared to FBLs, their results may not be conclusive.

In this work, in order to better understand the question ``why are some sources $\gamma$-ray detected and others are not ?", we compared various parameters of $Fermi$ BL Lacs (FBLs) with those of non-$Fermi$ BL Lacs (NFBLs) selected from cross-correlating the sample of 170 BL Lacs in Wu et al. (2007) with 2LAC AGN sample. The organization of this paper is as follows: the sample selection is presented in section 2; the comparisons of FBLs and NFBLs are shown in section 3; the discussions are given in section 4, and the results are summarized in section 5. Throughout the paper,
we define the spectral index $\alpha$ as $\rm
S_{\nu}\propto\nu^{-\alpha}$, where $S_{\nu}$ is the flux density at
frequency $\nu$, and a cosmology with $ H_{0}=70 \rm {~km ~s^
{-1}~Mpc^{-1}}$, $\rm \Omega_{M}=0.3$, $\rm \Omega_{\Lambda} = 0.7$ (e.g. Hinshaw et al. 2009) is adopted.




\section{The sample}

\cite{nieppola2006} presented a large sample of BL Lac objects, and the authors argued that this sample
is supposed to have no selection criteria (other than declination) in addition to
the ones in the original surveys. From this sample, \cite{wu06} estimated the
Doppler factor for a sample of 170 BL Lac objects using $ P_{\rm co5} $ = $ P_{\rm ci5}$ $\delta^{2+\alpha}$
and $\rm log~ \it P_{\rm ci5}$=$~0.62~\rm log~ \it P_{\rm t}+\rm 8.41$ derived for radio galaxies in Giovannini et al.
(2001), in which $ P_{\rm co5}$ is the observed
5 GHz core luminosity, $ P_{\rm ci5}$ is the intrinsic 5 GHz core luminosity, and $P_{\rm t}$ is the total radio luminosity at 408 MHz (see Wu et al. 2007, for details).

By cross-correlating the sample of 170 BL Lacs in \cite{wu06} with 2LAC, we define a subsample of 100 BL Lacs as FBLs sample, which were all detected with $Fermi$ LAT. The remaining 70 BL Lacs are included in NFBLs sample, as they were not detected with $Fermi$ LAT. \cite{linford12} selected samples of FBLs and NFBLs based on available VLBI images, in which the median values of the total VLBA flux density at 5 GHz are 177 mJy for FBLs and 221 mJy for 24 NFBLs, respectively. Compared to \cite{linford12} sample, the median flux density of our sample is much lower (see Table 1), especially for NFBLs, and the number of our NFBLs is about three times larger than theirs.

\section{The properties of FBLs and NFBLs }

To explore the differences of FBLs and NFBLs, we compare various properties for two subsamples, including the redshift, the low frequency radio luminosity at 408 MHz ($L_{\rm 408MHz}$), the absolute magnitude of host galaxies ($M_{\rm host}$), the polarization fraction from the NRAO VLA Sky Survey(NVSS) survey ($P_{\rm NVSS}$), the observed arcsecond scale radio core flux at 5 GHz ($F_{\rm core}$) and jet Doppler factor. The results are shown below.

\label{correlations}

\subsection{The distributions for $z$, $ L_{\rm 408MHz}$, $ M_{\rm host}$, $P_{\rm NVSS}$}
\label{correlations_int}
We compared several intrinsic parameters in FBLs with NFBLs, including redshift $z$, $L_{\rm 408MHz}$, $M_{\rm host}$, and $P_{\rm NVSS}$.
The redshift is available in 159 BL Lacs from NED\footnote{http://ned.ipac.caltech.edu/} and new measurements in \cite{shaw13}, including 81 FBLs, and 68 NFBLs. The $L_{\rm 408MHz}$ of all sources are directly adopted from Wu et al. (2007), which is the available arcsecond scale extended flux or was estimated from the radio flux at nearest frequency. The $L_{\rm 408MHz}$ is expected to be less influenced by beaming effect because it is comparable with $L_{\rm 408MHz}$ of FR I radio galaxies \citep{wu06}. $M_{\rm host}$ is available in \cite{wu09} for 121 BL Lacs, including 69 FBLs and 52 NFBLs. We collected the fraction of the polarized flux in total flux $P_{\rm NVSS}$ from NVSS, which is available for 92 FBLs and 55 NFBLs. The Kolmogorov Smirnov (KS) test shows that the distributions of FBLs and NFBLs do not show significant differences for all four parameters (see Table \ref{tablekstest}). However, the $\chi^2$-test does not show similarities between FBLs and NFBLs in the distributions of all four parameters.


\subsection{The Doppler factor and radio core flux} \label{dopandcore}
The Doppler factor $\delta$ can directly measure the significance of jet beaming effect, and the radio core flux usually are boosted by jet beaming effect in BL Lacs \citep{wu06}. These two parameters are available for all 170 BL Lacs in Wu et al. (2007). In Fig. \ref{dop_dis}, we compare the distributions of Doppler factor and the radio core flux $F_{\rm core}$ for all BL Lacs. We found that FBLs on average have larger Doppler factor $\delta$ and larger $F_{\rm core}$ than NFBLs. The KS test shows that there are significant differences between the total samples of FBLs and NFBLs in distributions of Doppler factor and $ F_{\rm core}$, all at the confidence level larger than 99.99\% (see Table \ref{tablekstest}). The systematically higher mean and median radio core flux in FBLs indicates that the intrinsic radio core flux may be higher than that in NFBLs at the fixed Doppler factor.


\begin{table*}

\caption{{\large The KS test on properties of \fermi and non-\fermi BL Lacs}\label{tablekstest}} \center
\begin{tabular}{lcccccccccccccc}
\hline\hline Parameter  & KS statistic & probability & Significantly different & subsets &$N_{\rm FBLs}$&$N_{\rm NFBLs}$&mean/median (FBLs)&mean/median(NFBLs)\\
\hline
$z$               &0.111  & 0.704  & NO  & allBLs       &  93&   66  &0.41/0.30& 0.35/0.30   \\
$\delta$              & 0.469 &1.29e-08   & YES  & allBLs               & 100   &70 & 5.8/3.7&2.9/2.1  \\
                & 0.412 &0.013   & YES  & LBLs               & 50   &19 &8.8/6.4&5.3/3.7\\
                & 0.583 &2.73e-04   & YES  & IBLs              & 24   &24 &3.4/3.3&2.1/2.0\\
               & 0.319 &0.107  & NO  & HBLs               & 26   &27 &2.5/2.2&1.9/1.9\\
Log $L_{\rm 408~ MHz}$ &  0.187 & 0.099   & NO  & allBLs   &  100   &70&25.6/25.5&25.1/25.1 \\
 $P_{\rm NVSS}$                & 0.146  &0.438  & NO  & allBLs              &   94    &53 &2.27/2.11&2.81/2.08\\
$ M_{\rm host}$              &0.210    & 0.132 & NO  & allBLs    &     71    & 50 &-22.9/-23.1&-23.0/-23.2         \\
$ F_{\rm core}$              &0.306   & 6.52e-04 & YES & allBLs    &     100    & 70 &0.55/0.23&0.10/0.023       \\
\hline
\end{tabular}
\end{table*}

 \begin{figure}
 \centering
 \includegraphics[height=5.5cm]{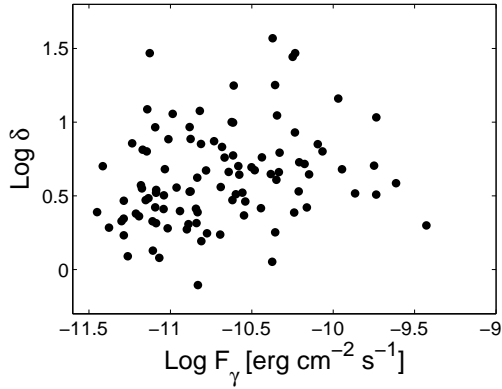}
 \caption{The $\gamma$-ray flux versus Doppler factor for FBLs.}
 \label{gammdetection}
 \end{figure}

\subsection{The $\gamma$-ray flux and Doppler factor}
\label{radiogamacor}
We collected the $\gamma$-ray flux $F_{\gamma}$ in the 100 MeV to 100 GeV range for all 100 FBLs from 2FGL. We show the relation between $F_{\gamma}$ and $\delta$ in Fig. \ref{gammdetection}. By using Spearman
rank correlation analysis \citep{mack82}, we found a significant correlation between the Doppler factor and $ F_{\gamma}$ with correlation coefficient $r=0.296$ at $>$ $99\%$ confidence level. A strong correlation(r=0.365 at $>$ $99.9\%$ confidence level) is also found between $\delta$ and $\gamma$-ray luminosity $ L_{\gamma}$. These results indicate that $\gamma$-ray emission is severely influenced by jet beaming effect. The Doppler factor is likely an important factor responsible for the detection of \fermi BL Lacs.





\section{Discussion}

\begin{figure}
\centering
\includegraphics[height=5.5cm]{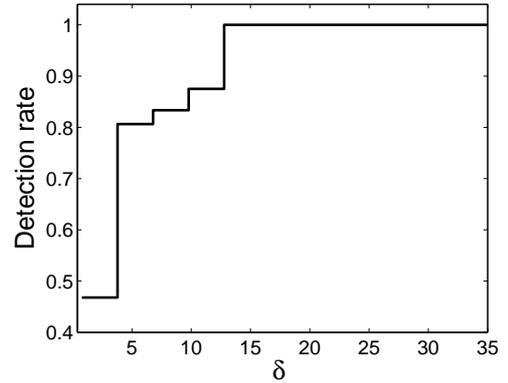}
\caption{The $Fermi~\gamma$-ray detection rate and Doppler factor.}
\label{detectionrate1}
\end{figure}

BL Lac objects are believed to be beamed FR I radio galaxies, and their multi-band electromagnetic radiation are related to the presence of relativistic particles in jets. Beaming is an important effect for understanding this type of objects. From a large sample of BL Lacs with estimated Doppler factor (Wu et al. 2007), we have found significant difference in Doppler factor of FBLs and NFBLs, while no significant differences in terms of several intrinsic parameters $z$, $L_{\rm 408 MHz}$, $M_{\rm host}$, and $P_{\rm NVSS}$. The Doppler factor of FBLs are systematically higher than those of NFBLs, implying important role of jet beaming effect in $\gamma$-ray detection. This systematic difference is also evident in the subsamples of LBLs and IBLs (see Table \ref{tablekstest}), but not in HBLs. However, the small number of sources precludes us to draw any firm conclusions in each subsamples. \cite{savolainen10}, using a sample consisting mostly of FSRQs and a different method of calculating the Doppler factor, reported a similar result in that Fermi-detected sources had larger Doppler factors, on average. In Fig. \ref{detectionrate1}, we plot the $Fermi~ \gamma$-ray detection rate (i.e. the ratio of the FBLs number to total BL Lacs in bins of Doppler factor, in which we use 3 as bin in Doppler factor) versus the Doppler factor. We found that the detection rate increases with Doppler factor when $\delta<10$, and it reaches 1.0 at $\delta>10$ and remains all the way to high $\delta$ values. This implies that sources with smaller Doppler factor may have smaller probability to have significant $\gamma$-ray emission.

\begin{figure}
\centering
\includegraphics[height=5.5cm]{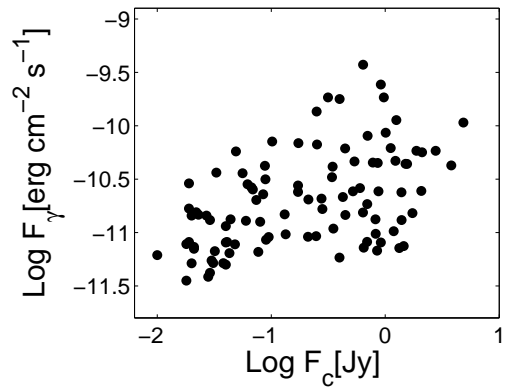}
\caption{The relation between radio core flux at 5 GHz and $\gamma$-ray flux in the 100 MeV to 100 GeV range. }
\label{radioandgamflux}
\end{figure}

\begin{figure}
\centering
\includegraphics[height=5.5cm, width=7.5cm]{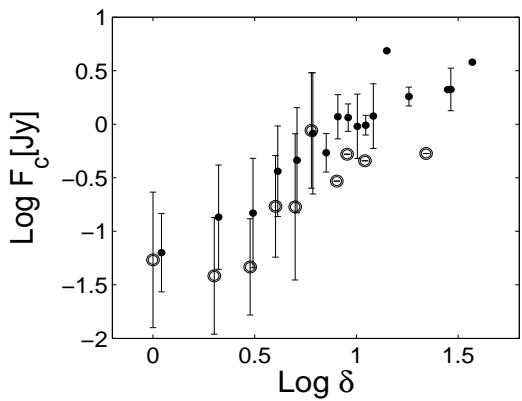}
\caption{ The relation between the Doppler factor and the average radio core flux in  binned $\delta$, with errorbars showing the standard deviation of core flux in each bin. The solid and open circles stand for FBLs and NFBLs, respectively.}
\label{fcoreanddelt}
\end{figure}
Proper motions can be used to study the bulk motion of jets when combining with other measurements. The superluminal motion have been often observed in blazars with relativistic jets moving towards us at small viewing angle. \cite{pin12} showed that sources in 2FGL display higher apparent speeds than those that are not detected by \fermi. We have searched literatures and found proper motion measurements for 38 sources, consisting of 33 FBLs and 5 NFBLs (see Table \ref{dataable}). Therefore, the measurements rate of proper motion are $\sim34\%$ and $\sim7\%$ for FBLs and NFBLs, respectively. Because the source selection for proper motion observations is usually not relevant with the information of $\gamma$-ray detection, so the available proper motion is hardly biased toward FBLs. The higher proper motion measurement rate in FBLs implies that the proper motion might be easier to be detected in FBLs, likely due to the higher jet beaming effect, thus consistent with higher Doppler factor in FBLs or the NFBLs are intrinsically dimmer.


\subsection{The radio core flux and $\gamma$-ray detection of BL Lacs}
Because $\gamma$-ray flux and radio core flux are Doppler boosted, a strong correlation between them is expected, which is plotted in Fig. \ref{radioandgamflux}. Indeed, a significant correlation is found with Spearman
correlation coefficient $r=0.493$ at $\gg99.99\%$ level. Interestingly, the correlation is still significant even after excluding the common dependence on the Doppler factor by using the partial Spearman correlation
method (Macklin 1982), with a correlation coefficient of 0.429 at $\gg99.99\%$ level. To further study the nature of NFBLs, we investigate the differences of FBLs and NFBLs in radio core flux at fixed Doppler factor. In Fig. \ref{fcoreanddelt}, we show the relationship between $\delta$ and the average $F_{\rm core}$ of FBLs and NFBLs in $\delta$ bins, in which we use 1 as the bin size in $\delta$. Intriguingly, FBLs have systematically larger radio core flux than NFBLs at fixed $\delta$, indicating larger intrinsic radio core flux in FBLs. By separating our 100 FBLs into two subsamples based on their $\gamma$-ray flux: 50 FBLs with relatively high $\gamma$-ray flux(HG-FBLs) (larger than 1.7E-11 $erg/cm^{2}/s$ )and others with relatively lower $\gamma$-ray(LG-FBLs), we also investigated the differences of HG-FBLs and LG-FBLs in radio core flux at fixed Doppler factor similar as we did for FBLs and NFBLs; the result is also similar with FBLs and NFBLs as showed in Fig. \ref{fcoreanddelt}, which means that HG-FBLs also have systematically larger radio core flux than LG-FBLs at fixed $\delta$, indicating larger intrinsic radio core flux in HG-FBLs. In combination with the correlation between $F_{\rm core}$ and $F_{\gamma}$, NFBLs may have relatively smaller
$F_{\gamma}$ even though they have comparable $\delta$ with FBLs, making them more difficult to be detected by \fermi LAT.

From Fig. \ref{fcoreanddelt}, it can be seen that the radio flux of NFBLs at fixed Doppler factor are systematically lower than FBLs, indicating that the $\gamma$-ray flux for NFBLs might also \textbf{be} lower, which likely results non-detection with \fermi telescope. Together with difference in $\delta$, the non-detection of $Fermi ~\gamma$-ray emission in NFBLs is likely due to their smaller Doppler factor and/or lower intrinsic $\gamma$-ray flux. However, it should be noticed that the radio flux of NFBLs covers a wide range overall or in each Doppler factor bin as shown by the large errorbar. The investigations on the differences between FBLs and NFBLs are further complicated due to the fact that BL Lac objects usually exhibit violent variations. Because the $\gamma$-ray and radio emission were not observed simultaneously, the variations thus may be at least partly responsible for the large dispersion of correlation between the $\gamma$-ray and radio flux (see Fig. 4). It is highly possible that NFBLs may enter a state where $\gamma$-ray production ceases or is at least significantly
reduced \citep{linford12}.
To exclude the effect of variations, 
 the simultaneous observations at $\gamma$-ray and other wavebands are required for larger samples of BL Lac objects, especially at radio bands.

There are two popular models for the $\gamma$-ray emission mechanism of blazars, the synchrotron self-compton model (SSC, seed photons from synchrotron radiation) and external-radiation-Compton (ERC, seed photons from external region) \citep{chen01}. Typically, ERC scenario requires that the inverse Compton (IC) radiation, i.e., $\gamma$-rays, originate relatively close to center, within central parsec, while SSC $\gamma$-rays come from jets beyond a parsec's distance \citep{nieppola2011}. Our results have shown that the non-detection of $\gamma$-ray is likely related with lower intrinsic radio core flux, which is originated from the jet synchrotron emission. The close realtions of $\gamma$-ray to radio emission may support the view of \cite{nieppola2011} that $\gamma$-rays are produced co-spatially with the arcsecond-scale radio core radiation, indicates that $\gamma$-ray is likely produced mainly through the SSC process in our sample.

\subsection{Effects of radio variability}
BL Lacs are known to show significant and rapid variations at nearly all wavebands. The effects of variations might be nontrivial in our results since the radio and $\gamma$-ray emission were not observed simultaneously. Although the $\gamma$-ray flux are variable, there should be little concern about short-term variability since the 2FGL data are basically averaged over $\sim$ 2 years. In this section, we only present the effects of radio variability on our statistical results.

\label{dandfcor}

The detailed variations on each source are basically unknown, we then have to make assumptions on the source state.  We considered 30\% or 50\% of sources were in elevated state with radio core flux or luminosity at 120$\%$ to 200$\%$ of the quiescent level. These 30\% or 50\% of sources were selected in three ways: (1) at highest radio flux; (2) at highest luminosity; (3) as random sources (see Table \ref{vartest1}). We recalculated the Doppler factor with newly assigned radio core quiescent flux or luminosity. The KS tests shows that there are still significant differences between FBLs and NFBLs in
distributions of Doppler factor and $F_{\rm core}$ in all three circumstances (see Table \ref{vartest1}). This strongly implies that the variability may not alter our results.

Alternatively, we investigated how much the variations are needed to eliminate the systematical difference between FBLs and NFBLs in $\delta$ and $F_{\rm core}$ distributions. As the $\delta$ and $F_{\rm core}$ for FBLs are on average much higher than NFBLs (see the mean and median values in Table \ref{tablekstest}),  we assume that the differences are totally due to the elevated state in FBLs during radio observation. Based on the mean values (see Table \ref{tablekstest}), we found that the significant differences disappear (the probability less than 90$\%$ from KS test), when we systematically reduce a factor of 1.7 - 1.9 for $\delta$, and a factor of 6 - 9 for $F_{\rm core}$ for FBLs, respectively. This means that the FBLs are expected to be in a flare state with flux density of a factor of 6 - 9 averagely higher than quiescent state, if the systematical differences between FBLs and NFBLs are completely caused by variations. However, the typical maximum variations are found to be around a factor of two, for invividual sources such as BL Lac \citep{villa2009}, and S5 0716+714 \citep{rani2013}, or from the light curves of BL Lac objects in University of Michigan Radio Astronomy Observatory (UMRAO) database\footnote{https://dept.astro.lsa.umich.edu/datasets/umrao.php}. Therefore, the systematically differences between FBLs and NFBLs cannot be caused by the variations only.

We further considered core flux variations within a factor of two in each sources, corresponding to $\delta$ variations within a factor of $\sqrt{2}$ \citep{gir04a}.  We let $\delta$ of each object to vary randomly between $\delta$$\times$$\sqrt{2}$ and $\delta/\sqrt{2}$ with a step size of 1\%. The KS tests on one million samples with allocated $\delta$ values, show that there is still a significant difference between FBLs and NFBLs in the $\delta$ distribution (confidence level larger than 99.8\%). By varying the core flux, a significant difference is also found between FBLs and NFBLs in the $F_{\rm core}$ distribution (confidence level larger than 99.9\%). We therefore are confident that the variations in the radio core flux will not significantly change our results.


We also tested the strong correlations of $\delta$-$F_{\gamma}$, $\delta$-$L_{\gamma}$ and $F_{\gamma}$-$F_{\rm core}$, by varying $F_{\rm core}$ randomly with a factor within [1/2, 2], and $\delta$ within [1/$\sqrt{2}$,$\sqrt{2}$]. On a million designed FBLs samples, we found that the significant correlations remain in almost all samples.

In summary, we argued that the variations in the radio core flux density will not affect our results, based on above various considerations.

\section{Summary}
By using the available data from literatures, we have compared various parameters of FBLs with those of NFBLs . We found that the Doppler factor is on average larger in FBLs than in NFBLs, and the $Fermi~ \gamma$-ray detection rate is higher in sources with higher Doppler factor.  The arcsecond scale radio core flux of NFBLs is on average lower than the FBLs at fixed doppler factor. Our results indicate that the Doppler factor is an important parameter of $\gamma$-ray detection. It seems that variations in the radio core flux density will not affect our results. The non-detection of $\gamma$-ray emission in NFBLs is likely due to low beaming effect, and/or low intrinsic $\gamma$-ray flux and the $\gamma$-rays seems to be produced co-spatially with the arcsecond-scale radio core radiation, and is likely produced mainly through the SSC process.

Because our sample is limited by the available archival data,
and the estimation of Doppler factor is based on an empirical
relation; the future new observational data including redshift, arcsecond scale radio data, and new improved estimations of Doppler factor for larger sample of BL Lac objects will all be useful for further tests of our results. Additionally, the new observations of BL Lac sample with less beaming effect using Very-long-baseline interferometry(VLBI) technique and at other wavelengths will be also important for understanding the nature of BL Lacs, and also helpful for further testing our results.

\begin{acknowledgements}
We thank the anonymous referee and editor for insightful comments
and constructive suggestions, which were greatly helpful in improving
our paper. We thank Emmanouil Papastergis for helpful discussions. This work is supported by the 973 Program (No. 2009CB824800), the NSFC grants (No. 11163002, 11073039, 11103060, 11233006, 11173043), and by the Study Abroad Fund from China Scholarship Council (No.[2011]5024).
\end{acknowledgements}

\Online
\begin{table*}

\caption{{\large The KS test for the variation effects on \fermi and non-\fermi BL Lacs}\label{vartest1}} \center
\begin{tabular}{lccc|ccccccccc}
\hline
Mode   &\multicolumn{3}{c|}{$\delta$ } & \multicolumn{3}{c}{$ F_{\rm core}$ }&\\

\hline   & KS statistic & probability & Significantly different & KS statistic & probability & Significantly different\\
\hline
$F120^{a}$              & 0.446 &8,0e-08   & YES  & 0.459          & 2.9e-08   &YES \\
$F150^{a}$               & 0.413 &8.7e-07  & YES  & 0.459              & 2.9e-08   &YES\\
$F200^{a}$               & 0.353 &4.5e-05   & YES  & 0.481             & 4.5e-09  &YES\\
$F120^{b}$              & 0.441 &1.1e-07   & YES  & 0.459          & 2.9e-08   &YES \\
$F150^{b}$               & 0.414 &7.9e-07  & YES  & 0.443              & 9.6e-08   &YES\\
$F200^{b}$               & 0.373 &1.3e-05   & YES  & 0.441             & 1.1e-07  &YES\\
$ L120^{a}$             & 0.450 &6.0e-08 &YES  & 0.471              & 1.0e-08   & YES \\
$ L150^{a} $            & 0.441 &1.1e-07  & YES  & 0.451               & 5.0e-08  &YES \\
$L200^{a} $            & 0.433 &2.0e-07  & YES  & 0.470               & 1.1e-08   & YES\\
$ L120^{b}$             & 0.453 &4.5e-08 &YES  & 0.467              & 1.4e-08   & YES \\
$ L150^{b} $            & 0.443 &9.6e-08  & YES  & 0.467               & 1.4e-08  &YES \\
$L200^{b} $            & 0.391 &3.8e-06  & YES  & 0.456               & 3.6e-08   & YES\\
$R120^{a} $            & 0.470 &1.1e-08  & YES  & 0.467               & 1.4e-08   &YES \\
$R150^{a} $             &0.477   & 6.4e-09 & YES & 0.487    &    2.8e-09    &YES        \\
$ R200^{a}$            & 0.453 &4.5e-08  & YES  & 0.487               &2.8e-09  &YES \\
$R120^{b} $            & 0.449 &6.2e-08  & YES  & 0.471               & 1.0e-08   &YES \\
$R150^{b} $             &0.436   & 1.7e-07 & YES & 0.461    &    2.9e-08    &YES        \\
$ R200^{b}$            & 0.401 &1.9e-06  & YES  & 0.464               &1.8e-08  &YES \\
\hline
\end{tabular}
\vskip 0.1 true cm \noindent
Notes: F120, F150, and F200 - a subsample of sources with the highest radio core flux, are assumed to be at elevated state with radio core flux at 120\%, 150\%, and 200\% of the quiescent state, respectively;  L120, L150, and L200: a subsample of sources with the highest radio core luminosity, are assumed to be at elevated state with radio core luminosity at 120\%, 150\%, and 200\% of the quiescent state, respectively; R120, R150, and R200: a subample of random sources are assumed to be at elevated state of 120\%, 150\%, and 200\% of the quiescent state, respectively. $^a$ and $^b$ indicate that the subsample size are 30\%, and 50\% of the whole sample, respectively.
\end{table*}

\begin{figure*}
\centering
\includegraphics[height=5.5cm, width=6.5cm]{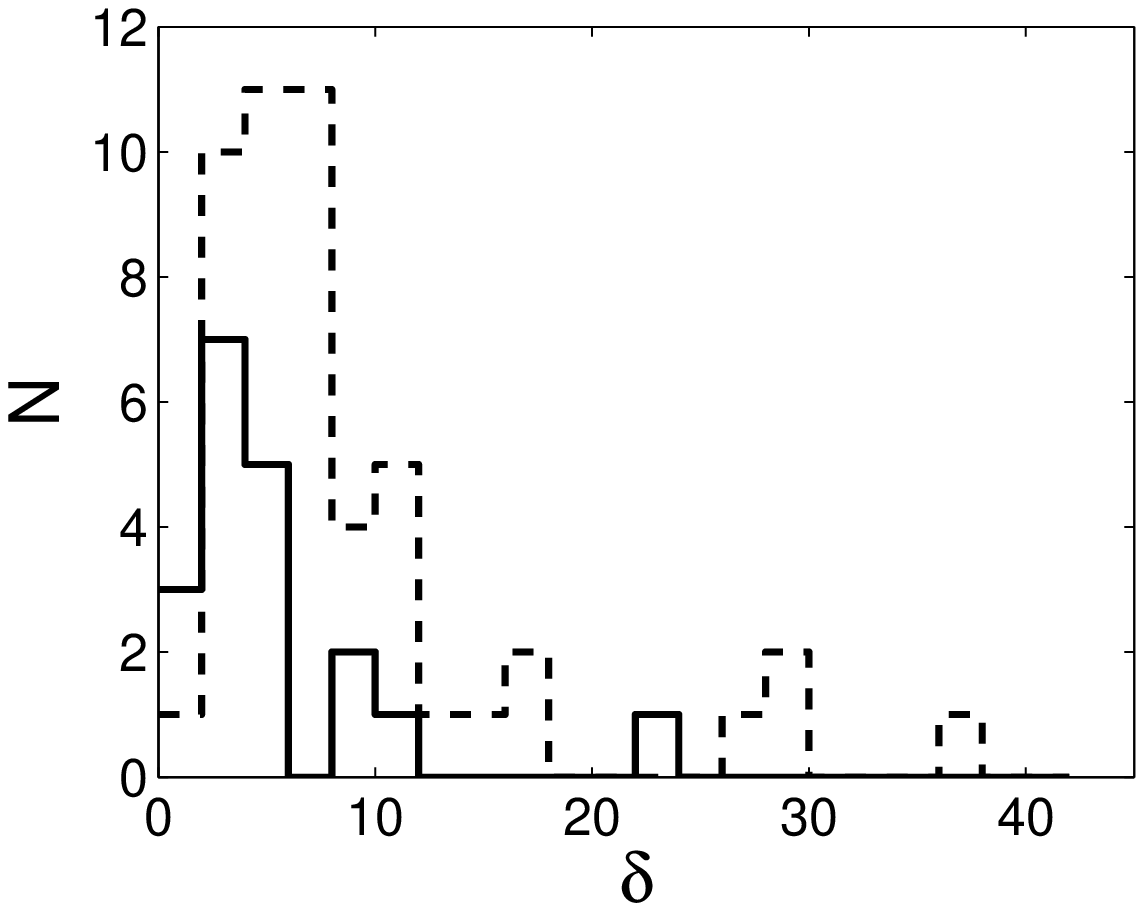}
\includegraphics[height=5.5cm, width=6.5cm]{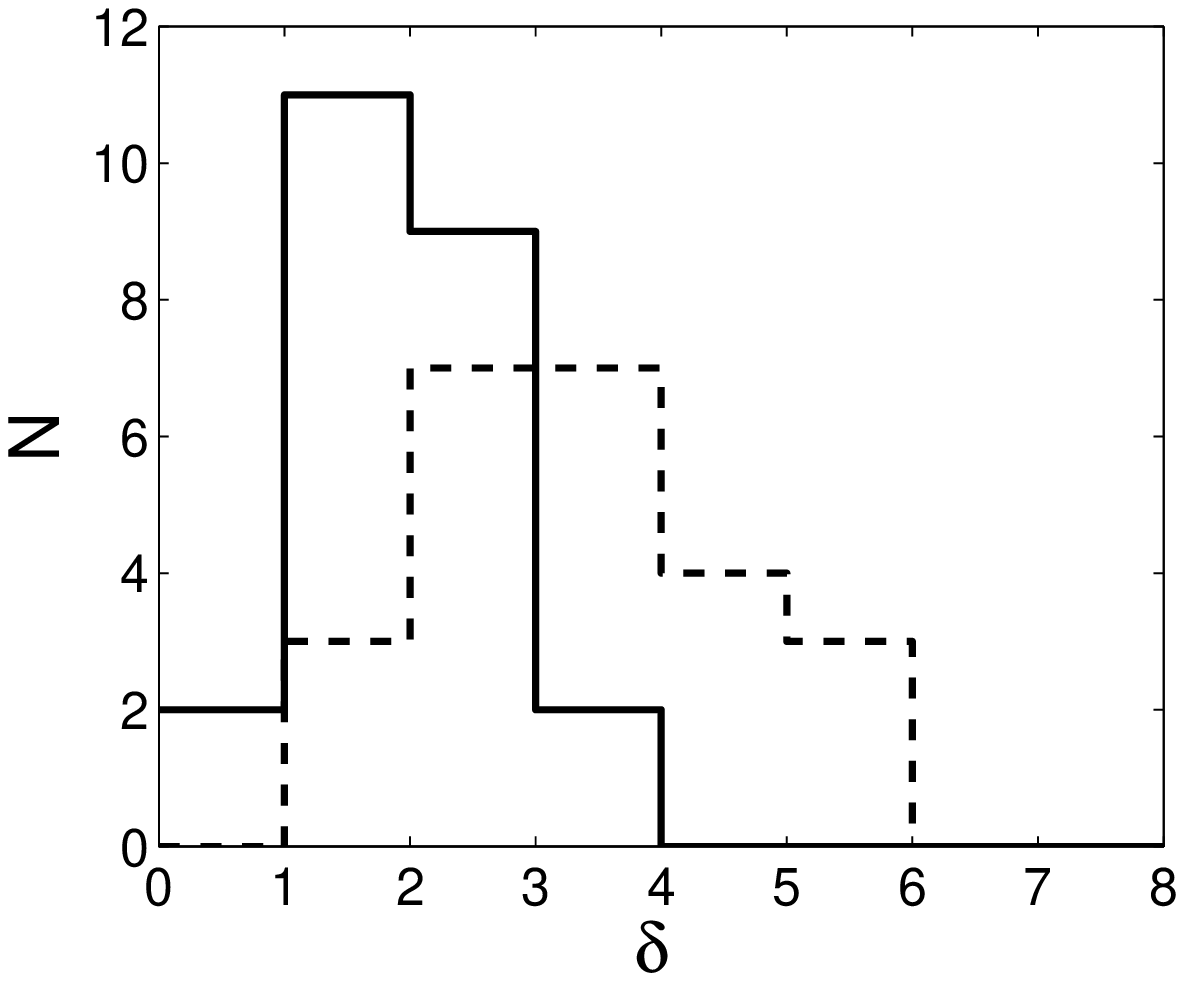}
\includegraphics[height=5.5cm, width=6.5cm]{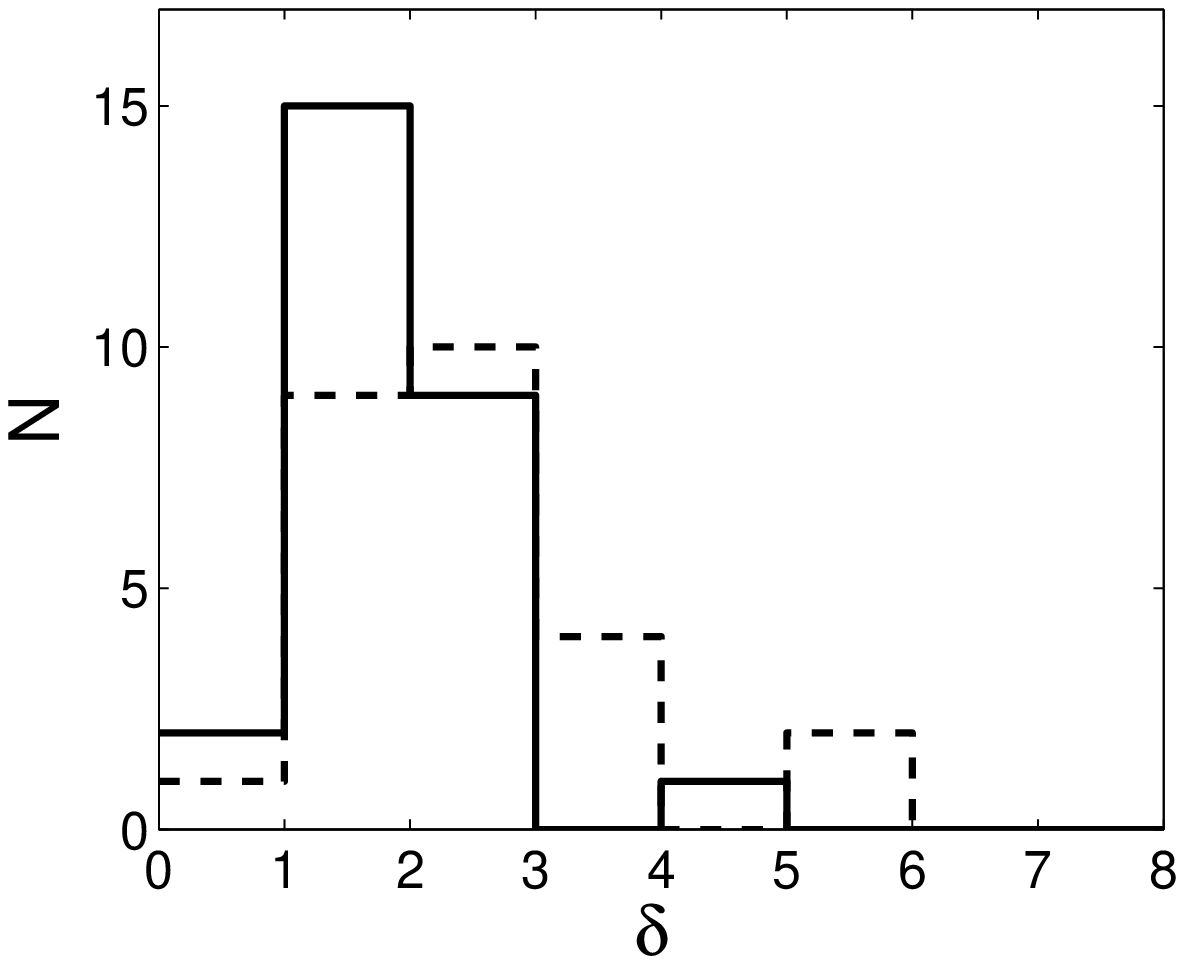}
\caption{The distributions of Doppler factor $\delta$ for LBL( top Left), IBL(top right) and HBLs(bottom). The dashed lines are for FBLs, while the solid lines for NFBLs.}
\label{dop_dislih}
\end{figure*}
\setlength{\tabcolsep}{0.05in}
\begin{longtable}{lllllllllllllll}
\caption{\label{dataable}The sample of 170 BL Lac objects in Wu et al. (2007).} \\
\toprule IAU name&Source&$z$ &$\rm \nu_{peak}^{'}$&log\,$P_{\rm 408~ M}$
  &$ F_{\rm core}$&$\delta$
&$P_{\rm NVSS}$& $M_{\rm host}$ &Class&$F_{\rm \gamma} $&$\beta_{\rm app}$&Ref. \\
&&&[Hz]&[Jy]&[Jy]&&&&&~$\rm [erg/cm^{2}/s]$&\\
(1)&(2)&(3)&(4)&(5)&(6)&(7)&(8)&(9)&(10)&(11)&(12)&\\
\midrule \midrule
\endfirsthead  
\midrule IAU name&Source&$z$ &$\rm \nu_{peak}^{'}$& log\,$P_{\rm 408~ M}$
  &$ F_{\rm core}$ &$\delta$
&$P_{\rm NVSS}$& $M_{\rm host}$ &Class&$F_{\rm \gamma}$&$\beta_{\rm app}$&Ref.\\
&&&[Hz]&[Jy]&[Jy]&&&&&~$\rm [erg/cm^{2}/s]$&\\
(1)&(2)&(3)&(4)&(5)&(6)&(7)&(8)&(9)&(10)&(11)&(12)&\\
\midrule
\endhead 

\midrule
\multicolumn{4}{r}{Continued\dots} \\
\endfoot 
\endlastfoot
0006-063  &  NRAO 5             &  0.347&     12.10&    26.94&      1.492   &   5.97&     2.69&      ...&    NFBL  &            ...&        2.89&          1               \\
0007+472  &  RX J0007.9+4711    &  2.100&     15.99&    27.68&    0.067     &   4.40&      1.08&  -27.08&    FBL   &           0.24E-10&        ...&    ...                   \\
0035+598  &  1ES 0033+595       &  0.086&     18.60&    24.01&     0.062     &  2.33&      0.66&  -20.41&    FBL   &           0.29E-10&       ...&     ...                  \\
0040+408  &  1ES 0037+405       &  0.271*&     16.62&    24.93&   0.015     &    1.92&     ...&       ...&   NFBL    &               ...&       ...&           ...            \\
0050-094  &  PKS 0048-097       &  0.635&     12.94&    27.47&     0.537     &  4.58&      3.13&     ...&    FBL   &          0.43E-10&        13.80&    2                     \\
0110+418  &  NPM1G +41.0022     &  0.096&     17.72&    24.41&     0.018     &  1.06&      3.64&  -22.94&    NFBL   &        ...&              ...&          ...             \\
0112+227  &  S2 0109+22         &  0.265&     12.84&    25.77&     0.700     &  7.09&      8.47&     ...&     FBL    &      0.79E-10&        ...&      ...                 \\
0115+253  &  RXS J0115.7+2519   &  0.350&     13.15&    25.31&     0.027     &  2.59&      ...&   -23.44&     FBL     &     0.14E-10&         ...&      ...                 \\
0123+343  &  1ES 0120+340       &  0.272&     17.96&    24.86&     0.031     &  2.93&      2.41&  -23.30&     FBL    &      0.58E-11&         ...&     ...                 \\
0124+093  &  MS 0122.1+0903     &  0.339&     15.36&    23.58&     0.001     &  1.96&      ...&   -23.07&     NFBL    &      ...&                    ...&           ...            \\
0136+391  &  B3 0133+388        &  0.271*&     16.31&    25.44&   0.049     &    2.44&     0.83&      ...&    FBL   &       0.62E-10&           ...&     ...                  \\
0141-094  &  PKS 0139-09        &  0.735&     12.77&    27.18&    0.696     &   7.43&     1.50&   -25.19&     FBL   &       0.18E-10&           5.50&    3                     \\
0148+140  &  1ES 0145+138       &  0.125&     15.76&    24.22&    0.002     &   0.54&     1.35&   -22.17&     NFBL   &       ...&             ...&          ...             \\
0153+712  &  8C 0149+710        &  0.022&     14.65&    23.99&    0.291     &   1.29&     3.10&   -23.23&     NFBL   &       ...&             ...&          ...             \\
0201+005  &  MS 0158.5+0019     &  0.299&     17.62&    24.47&    0.009     &   2.30&     1.97&   -23.11&     NFBL   &       ...&             ...&          ...             \\
0208+353  &  MS 0205.7+3509     &  0.318&     14.90&    23.91&    0.005     &   2.75&     8.30&      ...&     NFBL   &       ...&             ...&           ...            \\
0214+517  &  87GB 02109+5130    &  0.049&     17.53&    24.50&    0.161     &   1.50&     4.32&   -23.13&     NFBL   &       ...&             ...&          ...             \\
0222+430  &  3C 66A             &  0.440&     15.20&    27.55&    0.916     &   3.85&     1.75&      ...&       FBL   &    0.26E-09&        19.30&     4                    \\
0232+202  &  1ES 0229+200       &  0.140&     19.22&    24.75&    0.045     &   1.93&     2.15&   -23.77&      NFBL   &      ...&       ...&          ...             \\
0238+166  &  AO 0235+164        &  0.940&     12.82&    27.24&    0.972     &  10.78&     1.47&   -26.76&      FBL   &      0.18E-09&       25.60&    5                     \\
0301+346  &  MS 0257.9+3429     &  0.245&     13.10&    24.49&    0.009     &   1.90&     5.35&   -23.28&      NFBL   &      ...&       ...&          ...            \\
0314+247  &  RXS J0314.0+2445   &  0.054&     12.71&    22.95&    0.006     &   0.96&     1.52&   -21.26&      NFBL   &      ...&        ...&          ...             \\
0326+024  &  2E 0323+0214       &  0.147&     19.61&    24.16&    0.020     &   2.07&     3.03&   -22.63&      FBL    &     0.14E-10&         ...&     ...                  \\
0416+010  &  2E 0414+0057       &  0.287&     20.49&    25.70&    0.048     &   2.13&     2.22&   -24.02&      FBL    &     0.78E-11&         1.81&    6                     \\
0422+198  &  MS 0419.3+1943     &  0.512&     16.63&    25.15&    0.008     &   2.32&     3.64&   -23.54&         NFBL  &     ...&         ...&          ...             \\
0424+006  &  PKS 0422+004       &  0.268&     15.02&    26.19&    0.872     &   5.95&     2.18&      ...&       FBL    &     0.23E-10&         ...&      ...                 \\
0505+042  &  RXS J0505.5+0416   &  0.027&     16.87&    23.56&    0.090     &   1.20&     1.72&   -17.55&        FBL   &     0.85E-11&          ...&     ...                  \\
0507+676  &  1ES 0502+675       &  0.340&     18.55&    24.28&    0.033     &   5.77&     2.44&      ...&      FBL   &      0.42E-10&       ...&      ...                 \\
0508+845  &  S5 0454+84         &  1.340&     12.60&    26.28&    0.533     &  22.37&     0.30&   -22.89&       NFBL  &       ...&          ...&          ...             \\
0509-040  &  4U 0506-03         &  0.304&     17.77&    25.57&    0.029     &   1.93&     1.12&   -23.23&       NFBL  &       ...&          ...&          ...             \\
0613+711  &  MS 0607.9+7108     &  0.267&     14.41&    24.03&    0.014     &   3.49&     1.94&   -23.67&       NFBL  &       ...&          ...&          ...            \\
0625+446  &  87GB 06216+4441    &  0.311&     13.05&    25.83&    0.248     &   4.80&      4.70&     ...&      FBL   &       0.95E-11&        ...&     ...                  \\
0650+250  &  1ES 0647+250       &  0.203&     17.85&    24.85&    0.069     &   3.24&     0.25&   -21.39&      FBL   &       0.27E-10&        ...&     ...                  \\
0654+427  &  B3 0651+428        &  0.126&     14.80&    25.07&    0.134     &   2.38&     2.32&   -23.27&       NFBL  &       ...&           ...&          ...             \\
0656+426  &  NPM1G +42.0131     &  0.059&     17.34&    25.85&    0.253     &   0.86&     2.21&   -23.48&       NFBL  &       ...&           ...&          ...             \\
0710+591  &  EXO 0706.1+5913    &  0.125&     20.83&    25.03&    0.080     &   1.87&     0.32&   -23.27&     FBL    &       0.13E-10&        6.87&   7                      \\
0721+713  &  S5 0716+714        &  0.300&     14.06&    26.50&    0.315     &   3.23&     2.34&   -21.41&       FBL  &       0.18E-09&       14.80&    4                    \\
0738+177  &  PKS 0735+17        &  0.424&     12.64&    25.43&    2.775     &  29.41&     2.17&   -22.08&      FBL   &       0.57E-10&        7.40&    3                     \\
0744+745  &  MS 0737.9+7441     &  0.315&     13.24&    24.75&    0.021     &   3.05&     ...&    -23.53&      FBL   &       0.77E-11&       ...&      ...                 \\
0753+538  &  S4 0749+54         &  0.200&     12.23&    25.47&    1.390     &   9.26&     2.87&   -18.51&      FBL   &       0.13E-10&        ...&     ...                  \\
0757+099  &  PKS 0754+100       &  0.266&     12.48&    25.25&    2.073     &  17.72&     4.49&   -22.34&      FBL   &       0.23E-10&       14.40&    1                     \\
0806+595  &  SBS 0802+596       &  0.300&     16.43&    25.26&    0.029     &   2.36&     ...&    -24.32&       NFBL  &       ...&       ...&           ...            \\
0809+523  &  1ES 0806+524       &  0.137&     16.09&    24.90&    0.172     &   3.32&     3.04&   -23.22&      FBL   &       0.28E-10&        ...&     ...                  \\
0818+423  &  OJ 425             &  0.530&     12.71&    27.20&    1.011     &   6.33&     2.82&   -21.95&      FBL   &       0.81E-10&        4.90&    8                     \\
0823+223  &  4C 22.21           &  0.951&     12.94&    28.53&    0.388     &   2.73&     1.22&   -24.87&      NFBL   &       ...&       ...&          ...             \\
0825+031  &  PKS 0823+033       &  0.505&     11.79&    25.23&    1.453     &  29.41&     5.22&   -23.01&     FBL    &       0.76E-11&       17.80&    1                     \\
0831+044  &  PKS 0829+046       &  0.174&     12.81&    25.74&    1.230     &   6.21&     3.51&   -22.96&      FBL   &       0.47E-10&       10.10&    1                     \\
0831+087  &  1H 0827+089        &  0.941&     13.90&    26.67&    0.061     &   4.04&     0.27&      ...&       NFBL  &       ...&          ...&           ...            \\
0832+492  &  OJ 448             &  0.548&     12.33&    25.98&    0.294     &   8.39&     3.45&   -23.24&       NFBL  &       ...&          6.30&         3                \\
0854+441  &  US 1889            &  0.382&     17.31&    26.05&    0.031     &   1.79&     ...&       ...&       NFBL  &       ...&         ...&            ...           \\
0854+201  &  OJ 287             &  0.306&     12.75&    25.25&    1.557     &  17.87&     8.19&   -22.93&     FBL    &       0.41E-10&       15.17&    1                     \\
0915+295  &  B2 0912+29         &  0.302*&     15.64&    26.20&   0.172     &    2.96&     1.29&      ...&      FBL  &       0.25E-10&        ...&     ...                 \\
0916+526  &  RXS J0916.8+5238   &  0.190&     17.03&    25.26&    0.046     &   1.85&     0.86&   -23.88&       NFBL  &         ...&        ...&          ...             \\
0929+502  &  RXS J0929.2+5013   &  0.370&     13.76&    26.08&    0.916     &   9.23&     3.32&      ...&      FBL   &          0.81E-11&           ...&      ...                 \\
0930+498  &  1ES 0927+500       &  0.188&     20.77&    24.06&    0.018     &   2.70&     1.29&   -22.44&       NFBL  &         ...&        ...&          ...             \\
0930+350  &  B2 0927+35         &  0.302*&     14.35&    26.48&   0.394     &    3.68&     0.70&      ...&      NFBL  &         ...&        ...&          ...             \\
0952+656  &  RGB J0952+656      &  0.302*&     14.90&    25.46&   0.027     &    1.99&     4.95&      ...&      NFBL  &         ...&        ...&          ...             \\
0954+492  &  MS 0950.9+4929     &  0.380&     16.73&    24.20&    0.003     &   2.12&    11.15&      ...&       NFBL  &         ...&        ...&           ...            \\
0958+655  &  S4 0954+65         &  0.368&     13.06&    25.66&    0.276     &   6.79&     5.81&   -22.66&      FBL   &         0.18E-10&        6.20&    3                     \\
1012+424  &  RXS J1012.7+4229   &  0.365&     20.82&    25.83&    0.029     &   1.92&     0.61&   -23.84&      FBL   &         0.44E-11&        ...&     ...                  \\
1015+494  &  GB 1011+496        &  0.212&     16.40&    25.85&    0.173     &   2.64&     1.97&   -23.95&      FBL   &         0.73E-10&        ...&     ...                  \\
1031+508  &  1ES 1028+511       &  0.360&     18.16&    25.32&    0.044     &   3.39&     1.28&   -23.32&      FBL   &         0.14E-10&        ...&     ...                  \\
1037+571  &  RXS J1037.7+5711   &  0.830&     14.52&    26.48&    0.089     &   4.94&     0.65&      ...&      FBL   &      0.33E-10&       ...&      ...                 \\
1047+546  &  1ES 1044+549       &  0.540&     12.86&    24.82&    0.004     &   2.13&     ...&    -23.15&       NFBL  &      ...&        ...&           ...            \\
1053+494  &  MS 1050.7+4946     &  0.140&     14.95&    24.30&    0.040     &   2.49&     ...&    -23.98&      FBL   &      0.12E-10&       ...&      ...                 \\
1104+382  &  MRK 421            &  0.030&     18.20&    24.37&    0.639     &   1.99&     2.20&   -22.38&      FBL   &      0.38E-09&        0.80&    8                     \\
1109+241  &  1ES 1106+244       &  0.482&     16.69&    25.55&    0.018     &   2.45&     1.49&   -23.31&      FBL   &      0.40E-11&        ...&     ...                  \\
1120+422  &  EXO 1118.0+4228    &  0.124&     17.22&    24.10&    0.019     &   1.76&     1.91&      ...&      FBL   &      0.18E-10&         ...&      ...                 \\
1136+701  &  MRK 180            &  0.045&     18.74&    25.14&    0.131     &   0.78&     2.33&   -22.13&      FBL   &      0.15E-10&         2.23&    6                     \\
1136+676  &  RXS J1136.5+6737   &  0.134&     17.18&    24.16&    0.040     &   2.64&     3.64&   -23.24&      FBL   &      0.84E-11&         ...&     ...                  \\
1149+246  &  EXO 1449.9+2455    &  0.402&     19.63&    25.32&    0.015     &   2.20&     2.01&   -23.58&      NFBL   &      ...&          ...&          ...             \\
1150+242  &  B2 1147+245        &  0.200&     13.18&    25.30&    0.638     &   7.11&     1.91&   -19.56&     FBL    &      0.15E-10&         ...&     ...                 \\
1151+589  &  RXS J1151.4+5859   &  0.302*&     16.10&    25.98&   0.095     &    2.58&     1.05&      ...&     FBL   &      0.96E-11&          ...&     ...                  \\
1209+413  &  B3 1206+416        &  0.377&     13.87&    25.87&    0.397     &   7.19&     2.23&      ...&      FBL   &      0.67E-11&          ...&      ...                 \\
1215+075  &  1ES 1212+078       &  0.136&     15.57&    24.84&    0.091     &   2.49&     2.19&   -23.22&       NFBL  &      ...&           ...&          ...             \\
1217+301  &  B2 1215+30         &  0.130&     15.10&    25.46&    0.445     &   3.39&     5.38&   -23.15&      FBL   &      0.61E-10&          1.19&    6                     \\
1220+345  &  GB2 1217+348       &  0.643&     13.91&    26.64&    0.258     &   5.80&     1.16&      ...&       NFBL  &      ...&          ...&           ...            \\
1221+301  &  PG 1218+304        &  0.184&     18.80&    24.88&    0.056     &   2.61&     0.69&   -22.90&      FBL   &      0.38E-10&          ...&     ...                  \\
1221+282  &  ON 231             &  0.103&     14.15&    25.13&    1.117     &   5.34&     2.85&   -21.95&      FBL   &      0.62E-10&          3.20&    8                     \\
1223+806  &  S5 1221+80         &  0.430&     13.74&    26.89&    0.447     &   4.20&      1.45&     ...&      FBL   &      0.13E-10&          ...&     ...                  \\
1224+246  &  MS 1221.8+2452     &  0.218&     13.60&    24.25&    0.021     &   2.96&     2.08&   -21.84&      FBL   &      0.70E-11&          ...&     ...                  \\
1230+253  &  RXS J1230.2+2517   &  0.135&     14.40&    25.26&    0.351     &   3.60&     2.70&      ...&      FBL   &      0.11E-10&          ...&      ...                 \\
1231+642  &  MS 1229.2+6430     &  0.163&     15.84&    24.32&    0.042     &   2.96&     3.93&   -23.32&       NFBL  &      ...&           ...&          ...             \\
1237+629  &  MS 1235.4+6315     &  0.297&     15.61&    24.38&    0.014     &   3.02&     1.40&      ...&       NFBL  &      ...&           ...&           ...            \\
1241+066  &  1ES 1239+069       &  0.150&     17.05&    23.48&    0.010     &   2.45&     2.67&   -17.19&       NFBL  &      ...&           ...&          ...             \\
1248+583  &  PG 1246+586        &  0.847&     14.27&    26.90&    0.779     &  11.10&     2.23&   -24.23&     FBL    &      0.47E-10&          ...&     ...                  \\
1253+530  &  S4 1250+53         &  0.663&     14.39&    27.26&    0.346     &   4.44&      2.46&     ...&      FBL   &      0.42E-10&          ...&     ...                  \\
1257+242  &  1ES 1255+244       &  0.141&     16.83&    24.00&    0.007     &   1.30&     ...&    -22.59&      NFBL   &      ...&          ...&           ...            \\
1310+325  &  AUCVn              &  0.996&     12.61&    26.53&    2.107     &  27.73&     1.60&   -26.16&       FBL  &      0.52E-10&         20.88&    9                    \\
1322+081  &  1ES 1320+084N      &  0.049&     12.99&    22.74&    0.012     &   1.43&     ...&    -17.98&       NFBL  &      ...&           ...&           ...            \\
1341+399  &  RXS J1341.0+3959   &  0.169&     19.97&    25.22&    0.034     &   1.45&     ...&    -23.73&      NFBL   &      ...&           ...&           ...            \\
1402+159  &  MC 1400+162        &  0.244&     16.29&    26.54&    0.178     &   1.90&     6.03&      ...&       NFBL  &      ...&           ...&           ...            \\
1404+040  &  MS 1402.3+0416   &  0.344&     15.74&    25.76&    0.021     &   1.64&     ...&    -22.43&      NFBL   &      ...&           ...&           ...           \\
1409+596  &  MS 1407.9+5954     &  0.496&     16.40&    25.50&    0.017     &   2.55&     1.88&   -23.99&      NFBL   &      ...&            ...&          ...             \\
1415+485  &  RGB J1415+485      &  0.496&     14.13&    25.72&    0.058     &   4.06&     2.03&   -22.94&      NFBL   &      ...&            ...&          ...             \\
1415+133  &  PKS 1413+135       &  0.247&     12.39&    26.67&    0.694     &   3.47&     0.10&      ...&      FBL   &        0.72E-11&               7.10&     8                    \\
1417+257  &  2E 1415+2557       &  0.237&     19.00&    25.29&    0.040     &   2.13&     2.31&   -24.10&       FBL  &        0.50E-11&          ...&     ...                  \\
1419+543  &  OQ 530             &  0.152&     13.16&    24.67&    1.188     &  11.38&     1.25&   -23.40&       FBL  &        0.98E-11&          3.60&    3                     \\
1427+238  &  PKS 1424+240       &  0.300&     15.30&    26.31&    0.250     &   3.29&     1.12&   -20.29&       FBL  &        0.14E-09&          ...&     ...                  \\
1427+541  &  RGB J1427+541      &  0.106&     14.79&    24.38&    0.024     &   1.38&     ...&    -23.64&      NFBL   &        ...&           ...&           ...            \\
1428+426  &  H 1426+428         &  0.129&     18.41&    24.43&    0.022     &   1.56&     1.07&   -22.94&      FBL   &        0.17E-10&          ...&     ...                  \\
1439+395  &  PG 1437+398        &  0.260&     16.57&    25.70&    0.038     &   1.71&     1.73&   -23.12&      FBL   &        0.56E-11&          ...&     ...                  \\
1442+120  &  1ES 1440+122       &  0.163&     16.20&    24.81&    0.041     &   2.06&     1.32&   -23.01&       FBL  &        0.89E-11&          ...&     ...                  \\
1444+636  &  MS 1443.5+6349     &  0.298&     16.94&    24.31&    0.004     &   1.69&     4.66&      ...&      NFBL   &        ...&           ...&           ...            \\
1448+361  &  RXS J1448.0+3608   &  0.738&     16.44&    26.06&    0.029     &   3.38&      ...&      ...&      FBL   &        0.14E-10&           ...&      ...                 \\
1458+373  &  B3 1456+375        &  0.333&     12.86&    25.91&    0.305     &   5.38&     6.11&      ...&       NFBL  &        ...&             ...&           ...            \\
1501+226  &  MS 1458.8+2249     &  0.235&     14.69&    24.73&    0.085     &   4.59&     3.48&   -22.95&      FBL   &        0.24E-10&            ...&     ...                 \\
1509+559  &  SBS 1508+561       &  2.025&     15.00&    26.84&    0.028     &   5.03&     2.85&      ...&      FBL   &        0.50E-11&           ...&      ...                 \\
1516+293  &  RXS J1516.7+2918   &  0.130&     18.56&    25.01&    0.034     &   1.29&     2.57&   -23.41&       NFBL  &        ...&            ...&          ...             \\
1517+654  &  1H 1515+660        &  0.702&     17.82&    25.72&    0.019     &   3.32&     0.46&   -24.64&      FBL   &        0.98E-11&             ...&     ...                  \\
1532+302  &  RXS J1532.0+3016   &  0.064&     16.86&    23.87&    0.047     &   1.66&     2.08&   -22.43&      NFBL   &        ...&         ...&          ...             \\
1533+342  &  RXS J1533.4+3416   &  0.810&     17.89&    25.77&    0.033     &   4.86&     1.19&      ...&      NFBL   &        ...&          ...&           ...            \\
1534+372  &  RGB J1534+372      &  0.144&     13.97&    24.03&    0.020     &   2.22&     4.82&   -22.20&      FBL   &        \textbf{0.51E-11}&          ...&          ...             \\
1535+533  &  1ES 1533+535       &  0.890&     19.55&    25.97&    0.010     &   2.55&     4.44&   -26.35&      NFBL   &        ...&           ...&          ...             \\
1536+016  &  MS 1534.2+0148     &  0.312&     18.67&    25.53&    0.025     &   1.89&     1.65&   -23.43&      NFBL   &        ...&           ...&          ...             \\
1540+819  &  1ES 1544+820       &  0.271*&     17.53&    25.42&   0.043     &    2.30&     2.33&   -21.56&         FBL  &    0.73E-11&          ...&    ...                   \\
1540+147  &  4C 14.6            &  0.605&     14.25&    27.58&    1.329     &   6.34&     3.10&   -24.00&          FBL  &    \textbf{0.63E-11}&          8.73&         1                \\
1542+614  &  RXS J1542.9+6129   &  0.302*&     14.19&    25.27&   0.102     &    4.42&     1.33&      ...&         FBL  &    0.72E-10&          ...&     ...                  \\
1554+201  &  MS 1552.1+2020     &  0.222&     16.89&    25.09&    0.032     &   2.06&     3.04&   -23.77&          NFBL  &    ...&              ...&          ...            \\
1555+111  &  PG 1553+11         &  0.360&     15.92&    26.29&    0.398     &   5.07&     1.60&   -20.42&          FBL  &    0.20E-09&          ...&     ...                  \\
1602+308  &  RXS J1602.2+3050   &  1.091&     16.32&    26.70&    0.020     &   2.63&     1.78&      ...&          NFBL  &    ...&         ...&           ...            \\
1626+352  &  RXS J1626.4+3513   &  0.497&     15.05&    25.39&    0.014     &   2.53&     ...&    -24.36&          NFBL  &    ...&         ...&           ...            \\
1644+457  &  RXS J1644.2+4546   &  0.225&     17.28&    25.67&    0.064     &   1.95&     1.88&   -23.50&          NFBL  &    ...&           ...&          ...             \\
1652+403  &  RGB J1652+403      &  0.240&     14.80&    24.60&    0.011     &   1.85&     ...&       ...&          NFBL  &    ...&          0.54&           6              \\
1653+397  &  MRK 501            &  0.034&     16.17&    23.79&    1.251     &   4.79&     1.37&   -23.36&          FBL  &    0.11E-09&          1.50&    4                     \\
1704+716  &  RXS J1704.8+7138   &  0.350&     15.32&    25.06&    0.017     &   2.45&     2.06&   -22.83&          NFBL  &    ...&           ...&          ...             \\
1719+177  &  PKS 1717+177       &  0.137&     12.43&    25.19&    0.601     &   5.03&     3.70&      ...&          FBL  &    0.27E-10&         ...&      ...                 \\
1724+400  &  B2 1722+40         &  1.049&     12.64&    27.81&    0.341     &   4.72&     0.62&      ...&          FBL  &    0.26E-10&         ...&      ...                 \\
1725+118  &  H 1722+119         &  0.018&     15.76&    23.06&    0.088     &   1.13&     0.56&   -13.71&          FBL  &    0.42E-10&          ...&     ...                  \\
1728+502  &  IZw187             &  0.055&     17.16&    24.20&    0.134     &   1.91&     3.99&   -21.60&          FBL  &    0.97E-11&          5.30&    6                     \\
1739+476  &  OT 465             &  0.950&     13.39&    27.86&    0.848     &   6.50&     2.47&   -24.21&          FBL  &    0.63E-11&          ...&     ...                  \\
1742+597  &  RGBJ 1742+597      &  0.400&     13.75&    25.73&    0.077     &   3.73&     1.40&   -23.01&          FBL  &    0.61E-11&          ...&     ...                  \\
1743+195  &  NPM1G +19.0510     &  0.084&     17.44&    24.40&    0.210     &   3.19&     1.20&   -23.77&          FBL  &    0.94E-11&          1.17&    6                     \\
1745+398  &  B3 1743+398B       &  0.267&     17.58&    26.55&    0.118     &   1.69&     0.86&   -24.45&          NFBL  &    ...&            ...&          ...             \\
1747+469  &  B3 1746+470        &  1.484&     13.30&    27.26&    0.456     &  11.33&     1.87&      ...&          NFBL  &    ...&            ...&           ...            \\
1748+700  &  S4 1749+70         &  0.770&     13.80&    26.63&    0.521     &   9.94&     2.24&   -25.72&          FBL  &    0.24E-10&          3.20&    3                     \\
1749+433  &  B3 1747+433        &  0.571&     13.08&    26.82&    0.281     &   4.71&      5.82&     ...&          FBL  &    0.15E-10&          ...&     ...                  \\
1750+470  &  RXS J1750.0+4700   &  0.160&     18.31&    25.28&    0.010     &   0.72&     ...&    -23.09&          NFBL  &    ...&          ...&           ...            \\
1751+096  &  PKS 1749+096       &  0.322&     11.33&    24.92&    3.796     &  37.09&     2.00&   -23.20&          FBL  &    0.41E-10&          6.84&    9                    \\
1756+553  &  RXS J1756.2+5522   &  0.657&     19.74&    25.63&    0.010     &   2.40&      0.51&     ...&          FBL  &    0.68E-11&          ...&     ...                  \\
1757+705  &  MS 1757.7+7034     &  0.407&     13.37&    24.65&    0.011     &   3.02&     ...&    -22.80&          NFBL  &    ...&           ...&           ...            \\
1800+784  &  S5 1803+784        &  0.680&     13.35&    27.57&    1.878     &   8.51&     3.48&   -23.56&          FBL  &    0.52E-10&          8.97&    1                     \\
1806+698  &  3C 371             &  0.051&     14.42&    25.87&    1.508     &   1.79&     3.33&   -22.71&          FBL  &    0.44E-10&          2.90&    8                     \\
1808+468  &  RGB J1808+468      &  0.450&     14.34&    25.85&    0.041     &   2.80&     0.98&   -22.77&          NFBL  &    ...&           ...&          ...             \\
1811+442  &  RGB J1811+442      &  0.350&     15.30&    25.92&    0.006     &   0.79&     1.19&   -23.77&          NFBL  &    ...&           ...&          ...             \\
1813+317  &  B2 1811+31         &  0.117&     15.34&    25.00&    0.074     &   1.73&     2.48&   -21.02&          FBL  &    0.20E-10&          ...&     ...                  \\
1824+568  &  4C 56.27           &  0.664&     12.56&    28.02&    0.859     &   4.07&     1.56&   -24.13&          FBL  &    0.36E-10&         20.85&    1                     \\
1829+540  &  RXS J1829.4+5402   &  0.302*&     15.06&    25.73&   0.018     &    1.34&     1.57&      ...&         FBL  &    0.80E-11&          ...&     ...                  \\
1838+480  &  RXS J1838.7+4802   &  0.300&     13.24&    25.05&    0.023     &   2.45&     2.13&   -22.35&          FBL  &    0.16E-10&          ...&     ...                  \\
1841+591  &  RGB J1841+591      &  0.530&     14.94&    25.68&    0.006     &   1.44&     ...&    -24.33&          NFBL  &    ...&          ...&           ...            \\
1853+672  &  1ES 1853+671       &  0.212&     16.25&    23.81&    0.012     &   3.00&     6.79&   -22.27&          NFBL  &    ...&           ...&          ...             \\
1927+612  &  S4 1926+61         &  0.473*&     12.72&    26.62&   0.826     &    7.66&     2.42&      ...&         FBL  &    0.95E-11&          ...&     ...                  \\
1959+651  &  1ES 1959+650       &  0.047&     17.33&    23.01&    0.252     &   5.20&     1.42&   -22.20&          FBL  &    0.67E-10&          ...&     ...                  \\
2005+778  &  S5 2007+77         &  0.342&     12.39&    26.14&    0.822     &   7.70&     1.46&   -23.13&          FBL  &    0.12E-10&          2.90&    3                     \\
2009+724  &  S5 2010+72         &  1.740&     13.08&    28.41&    1.390     &  10.02&      1.35&     ...&          FBL  &    0.18E-10&          ...&     ...                  \\
2022+761  &  S5 2023+76         &  0.594&     13.54&    26.89&    0.425     &   5.75&      2.42&     ...&          FBL  &    0.19E-10&          ...&     ...                  \\
2039+523  &  1ES 2037+521       &  0.053&     15.60&    22.27&    0.032     &   3.55&     ...&    -23.22&          FBL  &    0.69E-11&         ...&      ...                 \\
2134-018  &  PKS 2131-021       &  1.285&     12.04&    27.93&    1.733     &  11.93&     1.19&   -25.83&          FBL  &    0.12E-10&          7.70&    8                     \\
2145+073  &  MS 2143.4+0704     &  0.237&     13.61&    25.14&    0.045     &   2.53&     1.21&   -22.96&          NFBL  &    ...&            ...&          ...             \\
2152+175  &  PKS 2149+17        &  0.874&     13.03&    26.67&    0.648     &  12.23&      1.06&  -23.11&          FBL  &    0.64E-11&             ...&    ...                   \\
2202+422  &  BL LAC             &  0.070&     13.15&    24.22&    4.857     &  14.47&     0.80&   -23.08&          FBL  &    0.11E-09&            6.50&    8                     \\
2250+384  &  B3 2247+381        &  0.119&     15.35&    24.65&    0.060     &   2.03&     3.03&   -23.21&          FBL  &    0.13E-10&            ...&     ...                  \\
2257+077  &  PKS 2254+074       &  0.190&     13.31&    24.78&    0.525     &   8.85&     4.89&   -23.61&          NFBL  &    ...&            4.30&         3                \\
2319+161  &  Q J2319+161        &  0.302*&     15.30&    25.14&   0.017     &    2.00&     3.03&      ...&         NFBL  &    ...&            ...&          ...             \\
2322+346  &  TEX 2320+343       &  0.098&     16.68&    24.58&    0.030     &   1.23&     ...&    -23.39&          FBL  &    0.56E-11&           ...&      ...                 \\
2323+421  &  1ES 2321+419       &  0.268&     12.90&    24.52&    0.019     &   2.90&     2.57&      ...&          FBL  &    0.30E-10&           ...&      ...                 \\
2329+177  &  1ES 2326+174       &  0.213&     17.84&    24.64&    0.018     &   2.04&     3.03&   -22.94&          NFBL  &    ...&             ...&          ...             \\
2339+055  &  MS 2336.5+0517     &  0.740&     14.89&    25.74&    0.005     &   1.82&     ...&       ...&          NFBL  &    ...&             ...&            ...          \\
2347+517  &  1ES 2344+514       &  0.044&     15.86&    23.30&    0.212     &   3.63&     0.62&   -23.06&          FBL  &    0.21E-10&            1.15&    6                     \\
2350+196  &  MS 2347.4+1924     &  0.515&     15.72&    24.80&    0.003     &   1.81&     ...&       ...&          NFBL  &    ...&             ...&            ...           \\

\bottomrule

\end{longtable}
\vspace{-4mm} Notes: Col.4-7 were first reported in \cite{wu06}, Col. 9 was first reported in \cite{wu09} and these values have been revised using redshift from Col 3. The columns are: Col. 1: the source IAU name (J2000). Col. 2:
the source alias name. Col. 3: the redshift. `*' indicate that the
redshift is unknown, and taken as the average redshift of
LBL/IBL/HBL subclass.  Col. 4: the synchrotron peak frequency. Col. 5: the total
radio power at 408 MHz. Col. 6: The arcsecond scale radio flux at 5 GHz. Col. 7: the Doppler factor. Col. 8: the polarization fraction from NVSS in per cent. Col. 9: Absolute Magnitude of host galaxies. Col. 10, Indication of FBLs and NFBLs. Col. 11, observed $\gamma$-ray flux from 2nd year \fermilat catalogue[100 MeV to 100 GeV]. Col. 12: the proper motion from literatures. Col. 13, the references of proper motion:  1.
\cite{kharb10}
 2.     \cite{pin12}
 3.    \cite{Gabuzda00}
 4.     \cite{jorstad01}
 5.    \cite{pin06}
 6.     \cite{kharb08}
 7.     \cite{wu12}
 8.     \cite{kellermann04}
 9.    \cite{lister09}

\clearpage


\begin{thebibliography}{}
\bibitem[Abdo et al.(2009)]{abdo2009}
Abdo, A. A., Ackermann, M.,  Ajello, M., et al. 2009, ApJ, 700, 597

\bibitem[Abdo et al.(2010)]{abdo2010}
Abdo, A. A., Ackermann, M.,  Ajello, M., et al. 2010, ApJ, 715, 429

\bibitem[Ackermann et al.(2011)]{ackermann2011}
Ackermann, M., Ajello, M., Allafort, A., et al. 2011, ApJ, 743, 171



\bibitem[Atwood et al.(2009)]{Atwood09}
Atwood, W. B., Abdo, A. A., Ackermann, M., et al. 2009, ApJ, 697, 1071

\bibitem[Chen \& Bai (2011)]{chen01}
Chen, L., \& Bai, J. M., 2011, \apj, 735, 108


\bibitem[Gabuzda et al.(2000)]{Gabuzda00}
Gabuzda, D. C., Pushkarev, A. B., Cawthorne, T. V., 2000, MNRAS,
319, 1109

\bibitem[Giovannini et al.(2001)]{giova01}
Giovannini, G., Cotton, W. D., Feretti, L., Lara, L., \& Venturi, T.
2001, \apj, 552, 508

\bibitem[Hinshaw et al.(2009)]{hinshaw09}
Hinshaw, G., Weiland, J. L., Hill, R. S. et al. 2009, ApJS, 180, 225



\bibitem[Giroletti et al.(2004)]{gir04a}
Giroletti, M., Giovannini, G., Taylor, G.~B., \& Falomo, R. \ 2004,
ApJ, 613, 752








\bibitem[Jorstad et al.(2001)]{jorstad01}
Jorstad, S. G., Marscher, A. P., Mattox, J. R., et al. 2001, \apj, 134, 181


\bibitem[Kellermann et al.(2004)]{kellermann04}
Kellermann, K. I., Lister, M. L., Homan, D. C., et al. 2004, \apj, 609, 539

\bibitem[Kharb et al.(2008)]{kharb08}
Kharb, P., Gabuzda, D., Shastri, P. et al. 2008, MNRAS, 384, 230

\bibitem[Kharb et al.(2010)]{kharb10}
Kharb, P., Lister, M. L., Cooper, N. J., 2010, \apj, 710,764

\bibitem[Linford et al.(2011)]{linford11}
Linford, J. D., Taylor, G. B., Romani, R., et al. 2011, ApJ, 726, 16

\bibitem[Linford et al.(2012)]{linford12}
Linford, J. D., Taylor, G. B., Romani, R., et al. 2012, ApJ, 744, 177


\bibitem[Lister et al.(2009)]{lister09}
Lister, M. L., Homan, D. C., Kadler, M., 2009, \apj, 696, 22


\bibitem[Macklin (1982)]{mack82}
Macklin, J.T. 1982, MNRAS, 199, 1119

\bibitem[Nieppola et al.(2006)]{nieppola2006}
Nieppola, E., Tornikoski, M., \& Valtaoja, E. 2006, A\&A, 445, 441

\bibitem[Nieppola et al.(2011)]{nieppola2011}
Nieppola, E., Tornikoski, M., \& Valtaoja, E. 2011, A\&A, 535, 69

\bibitem[Nolan et al.(2012)]{nolan12}
Nolan, P. L., et al. 2012, ApJS, 199, 31

\bibitem[Padovani \& Giommi (1995)]{padovani95}
Padovani, P., \& Giommi, P. 1995, ApJ, 446, 547




\bibitem[Piner et al. (2006)]{pin06}
Piner, B. G., Bhattarai, D., Edwards, P. G., Jones, D. L., 2006, \apj, 640, 196

\bibitem[Piner et al. (2012)]{pin12}
Piner, B. G., Pushkarev, A. B., Kovalev, Y. Y., et al., 2012, ApJ, 758, 84p


\bibitem[Pushkarev et al. (2012)]{pushkarev12aa}
Pushkarev, A. B., Kovalev, Y. Y., Lister, M. L., Savolainen, T. 2012, A\&A, 544, 3p

\bibitem[Rani et al. (2013)]{rani2013}
Rani, B., Krichbaum, T. P., Fuhrmann, L., et al. 2013, A\&A, 552,11R



\bibitem[Savolainen et al.(2010)]{savolainen10}
Savolainen, T.,  Homan, D. C.,  Hovatta, T., et al., 2010, A\&A, 512, A24
\bibitem[Shaw et al.(2013)]{shaw13}
Shaw, M. S., Romani, R. W., Cotter, G., et al. 2013, \apj, 764, 135S

\bibitem[Urry \& Padovani (1995)]{urry95}
Urry C.~M., Padovani P., 1995, \pasp, 107 , 803

%
\bibitem[Villata et al. (2009)]{villa2009}
Villata, M.; Raiteri, C. M.; Larionov, V. M.;  et al. 2009, A\&A, 501, 455

\bibitem[Wu et al.(2007)]{wu06}
Wu, Z. Z., Jiang, D., R., Gu, M. F., Liu, Y., 2007, A\&A, 466, 63

\bibitem[Wu et al.(2009)]{wu09}
Wu, Z. Z.,  Gu, M. F., Jiang, D., R., 2009, RAA, 9, 168


\bibitem[Wu et al.(2012)]{wu12}
Wu, Z. Z.,  Gu, M. F., Jiang, D., R., 2012, MNRAS, 424, 2733

\end{thebibliography}
\end{document}